# REMEMBER WHEN? DECIPHERING EDIACARAN-CAMBRIAN METAZOAN BEHAVIOUR AND MEMORY USING FOSSIL MOVEMENT PATHS

Brittany A. Laing.[1,2], M. Gabriela Mángano[1], Luis A. Buatois[1],Glenn A. Brock[2], Romain Gougeon[1], Zoe Vestrum[3], Luke C. Strotz[4], & Lyndon Koens[5]

## 4.1 ABSTRACT

Evaluating the timing and trajectory of sensory system innovations is crucial for understanding the increase in phylogenetic, behavioural, and ecological diversity during the Ediacaran-Cambrian transition. Elucidation of sensory adaptations has relied on either body-fossil evidence based on anatomical features or qualitative descriptions of trace-fossil morphology, leaving a gap in the record of sensory system innovations between the development of basic sensory capacities and that of more advanced sensory organs and brains. Here, we examine fossil movement trajectories of Ediacaran and Cambrian grazers for the presence of autocorrelation. Our analysis reveals a lack of temporal correlation in the studied Ediacaran trajectories and its presence in both analysed Cambrian trajectories, indicating time-tuned behaviours were in place by the early Cambrian. These results support the Cambrian Information Revolution hypothesis and indicates that increases in cognitive complexity and behavioural strategies were yet another important evolutionary innovation that occurred during the Ediacaran Cambrian transition.

[1] *Department of Geological Sciences, University of Saskatchewan, Saskatoon, Saskatchewan, Canada.*
[2] *School of Natural Sciences, Macquarie University, Sydney, New South Wales, Australia.*
[3] *Department of Physics, University of Alberta, Edmonton, Alberta, Canada.*
[4] *State Key Laboratory of Continental Dynamics, Shaanxi Key Laboratory of Early Life & Environments and Department of Geology, Northwest University, Xi'an, China.*
[5] *Department of Mathematics, University of Hull, United Kingdom.*

4.2 INTRODUCTION

The ability to sense and respond to external signals is crucial to an organism's survival. The timing, ecological context, and trajectory of the evolution of sensory systems are accordingly topics of great interest to evolutionary biologists, macroecologists, and paleontologists. The complexity and size of these systems varies widely across organisms, from those contained within a single cell, to vast somatosensory systems with dedicated sensory organs and processing regions (Niven & Laughlin, 2008 and sources therein). Sensory and cognitive complexity is thought to be predominantly impacted by the cost-benefit ratio of two selective pressures; the need to obtain more information from the external environment, and the energy cost incurred by increasingly complex sensory systems (Niven & Laughlin, 2008; Plotnick et al., 2010).

One hypothesis, the Cambrian Information Revolution, suggests a coevolutionary increase of these selective pressures occurred during the Ediacaran-Cambrian transition (Plotnick et al., 2010). During this time an increase in phylogenetic diversity was accompanied by a first-order transition in ecological structuring, from mat-ground dominated ecosystems with limited infaunal and pelagic activity to a truly Phanerozoic ecosystem with increased exploitation of infaunal and pelagic realms along with a rise of more diverse feeding styles and motility levels (Seilacher & Pflüger, 1994; Seilacher, 1999; Droser et al., 2017; Mángano & Buatois, 2017; Wood et al., 2019; Mángano & Buatois, 2020). As a result, organism would have to contend with increasing spatial complexity and a more diverse signal landscape (i.e. increasing amounts of information important to a species survival). The Cambrian Information Revolution correlates this increasingly information-rich environment to the evolution of sensory organs and innovative behaviours during the Ediacaran-Cambrian transition (Plotnick et al., 2010). In the information-poor Ediacaran, sensory systems would have likely imposed an unnecessary cost. Yet in the Cambrian, with the increase in ecological complexity associated with the Cambrian radiation (Bush & Bambach, 2011), the ability to better detect prey, predators, and resources would have become paramount to species survival. Insights from molecular clock and developmental regulatory gene research (i.e deep homologies) indicate a deep root for the evolution of sensory structures (Jacobs et al., 2007; Shubin et al., 2009), suggesting the foundation for sensory systems and organs were well in place by the Ediacaran (Hsieh et al., 2022). Spatial statistical methods (Paterson et al., 2017; Coutts et al., 2018) and trace-fossil analysis (Carbone & Narbonne, 2014; Gehling & Droser, 2018) have provided insights on the sensory capabilities of some Ediacaran organisms. In turn, all extant organisms which are capable of directed self-propelled movement possess, at minimum, a basic sensory capacity (e.g. chemoreception, odor-gated rheotaxis) (Hildebrand, 1995). It is likely, then, that the makers of early trace fossils, unequivocally present since the late Ediacaran (~ 560 Ma) (Jensen et al., 2007; Wood et al., 2019; Mángano & Buatois, 2020), also had a basic ability to sense and respond to external signals (Carbone & Narbonne, 2014; Hsieh et al., 2022). While visual sensory organs and brains can be found near the end of the early Cambrian (~520 Ma) (Zhang et al., 1990; Schoenemann, 2006; Clarkson et al., 2006; Schoenemann et al., 2009; Ma et al., 2012; Zhao et al., 2013; Cong et al., 2014; Ma et al., 2015; Edgecombe et al., 2015; Strausfeld et al., 2016a; 2016b; Paterson et al., 2020; Hsieh et al., 2022), discernable sensory organs are unreported from the Ediacaran (Evans et al., 2021). This leaves a notable gap (~30 Ma) in the evolutionary record of sensory systems, between the evolution of basic sensory capacities (e.g. chemoreception, nerve nets) and the development of complex sensory and associated neurological systems with dedicated sense organs (e.g. brains, eyes).

Fossil movement trajectories offer a prolific data source for investigating this gap. The Movement Ecology Paradigm (Nathan et al., 2008) offers a way to examine the formation of these trajectories methodologically. It describes the formation of a movement path as a function of four factors: (1) the organism's intrinsic motivation to move (i.e. internal state), (2) the organism's ability to sense and respond to external signals (i.e. navigation capacity), (3) the organism's basic ability to move (i.e. motion capacity), and (4) the external factors which affect movement, such as substrate, food distribution, predation, and competition (i.e. external factors). As the expression of these factors change, so will the resulting trajectory. Consequently, trajectories can be recorded and analysed to hypothesize on the expression of these factors.

The potential of fossil trajectories to provide insight on the evolution of navigation capacities has long been known (Seilacher, 1967). Of particular focus is the evolutionary trajectory of grazing behaviour during the Ediacaran-Cambrian transition, due in part to the inferred impact of the eventual restriction of Ediacaran mat-grounds. The appearance of morphologically regular trace-fossils in the Cambrian, such as *Psammichnites saltensis*, has long been suggested to reflect more sophisticated behaviours and sensory adaptations, which allowed for more efficient exploitation of nutrient-rich zones (Seilacher, 1967; 1977; 1997; Crimes, 1992; Kim, 1996; Crimes & Droser, 1992; Carbone & Narbonne 2014; Mángano & Buatois, 2016). These inferred strategies have relied on qualitative descriptions of path morphology (e.g. irregular, circular scribbles, looping, first or second order meanders) to make inferences on sensory adaptations and navigation capacities (though see Fan et al. 2017). Traditionally, these inferences reference the optimization of foraging behaviours (Seilacher, 1967), with increasingly regular paths indicating increasing cognitive complexity and adaptation to heterogenous environments (e.g. irregular paths, to more efficient scribbling or circling paths, to meandering paths). The reliance on qualitative descriptors results in a difficulty to identify the appearance and trajectory of specific innovations in sensory and navigation capacities during the Ediacaran-Cambrian transition. Thus, our understanding of the timing, ecological context, and trajectory of the evolution of sensory systems and behavioural strategies during this critical evolutionary period remains incomplete.

We propose the use of quantitative techniques to examine and assess movement trajectories inferred from trace fossils produced by purported metazoans across the Ediacaran-Cambrian transition for temporal trends. Similar analyses of extant organism movement trajectories have demonstrated their utility to detect temporal and spatial patterns in animal movement and relate this rhythmicity to internal states, search strategies, environmental heterogeneity, and the presence of memory-driven directed behaviour (Austin et al., 2004; Wittenmyer et al., 2008; Boyce et al., 2010; Thompson et al., 2022).

Here, we apply a newly developed method (**Chapter 3**) for discretizing fossil movement paths to analyse Ediacaran and Cambrian trajectories for temporal trends. Specifically, we will be examining trajectories for the presence of temporally correlated turning angles to reveal any presence of temporally repeating patterns. To do so, we ran autocorrelation and partial autocorrelation analyses to examine if turns are correlated to turns taken a certain time prior. In mathematics this time-series correlation is referred to as "memory", as future values are dependent on past ones. This is different to the definition of memory employed by biologists, where memory refers to the ability of an organism to acquire, encode, store, and retrieve information (Baddeley, 2004; Fagan et al., 2013).

The presence of temporal autocorrelation does not always indicate the presence of time-tuned behaviours or biologic memory, as other factors could account for autocorrelation (Dray et al., 2010; Fagan et al., 2013). Such patterns in the turning angles of fossil movement trajectories could be the product of anatomical constraints on motion (i.e. the motion capacity), such as the gait of a swimming stroke. However, this would only account for a single time-series pattern (i.e. an isolated Fourier mode) and would exist only for times shorter than or equal to the time taken to complete a single swimming stroke. In turn, only those external factors with strong periodicities (e.g. light levels related to solar cycles, water level related to tidal cycles, etc.) could cause temporally correlated behaviour. It is highly unlikely that responses to less periodic external factors, such as nutrient distribution or sediment consistency, could account for the mathematically repeating locomotory patterns revealed by our autocorrelation analyses. As a result, we interpret any autocorrelation at times larger than the time taken to complete a single action or swimming stroke to be internally driven (i.e. internal state) and evidence of a time-tuned behaviour or possible biologic memory (i.e. navigation capacity). Our analysis thus seeks to determine if the turning angles of fossil trajectories are autocorrelated, inferred to indicate the presence of a time-tuned behavioural strategy or possible biologic memory.

In environments interpreted to have complex information landscapes, we expect this alternate hypothesis to be true, while in environments interpreted to have simple information landscapes, we expect the null hypothesis (that fossil trajectories are uncorrelated) to be true (Plotnick et al., 2010). This results in two predictions. First, that the trajectories of Ediacaran specimens will be uncorrelated. With a greater abundance of microbial mats and a less complex signal landscapes, the advantage incurred by time-tuned behaviours would be minimal in comparison to the energy cost imposed by this sensory adaptation (Droser et al., 2017; Tarhan et al., 2022). The second prediction would be that the trajectories of Cambrian specimens will be autocorrelated. As the environment became increasingly information-rich, with increasingly patchy distribution of food sources, the advantage provided by a time-tuned behaviours would outweigh the energy cost imposed by this navigation capacity adaptation.

4.3 METHODS

To examine the navigational capacity of organisms across the Ediacaran-Cambrian transition our investigation focuses on specimens which record horizontal locomotion with a purported feeding aspect (i.e. pascichnial trails). As a result, all our specimens consist of bedding-parallel trails with, save for *Helminthoidichnites tenuis*, active infill. We chose ensembles of trace fossil specimens which could be reasonably grouped as a single population, that is, specimens which were located on a few large stratigraphic surfaces (bedding planes), and which can be reasonably assumed to reflect the activity of a single species of tracemaker (i.e. one ichnospecies of similar trail widths per locality). The four ichnospecies include two from the Ediacaran and two from the Cambrian.

Our oldest specimens are classified as *Helminthoidichnites tenuis* and come from the Ediacara Member (Rawnsley Quartzite) of Australia, which consists largely of sandstones deposited in a shallow marine environment (Gehling, 2000; Gehling & Droser, 2012; McMahon et al., 2020, Figures S1 & S2). Overlap of associated taxa with assemblages in the White Sea region of Russia provide a tentative age between 560 and 551 Ma (Evans et al., 2021 and sources therein). *Helminthoidichnites tenuis* is commonly described as a random, unordered path (e.g. Hofmann, 1990). Recent work on specimens from the Ediacara Member (Australia) suggests the tracemaker deflected towards the bodies of megascopic Ediacaran organisms (Gehling & Droser, 2018), which would require a basic sensory modality (e.g. chemoreception). The second set of Ediacaran trajectories come from specimens of *Parasammichnites pretzeliformis* from the Spitskop Member (Urusis Formation) of Namibia. Fossils in these strata are representative of the Nama assemblage (549-539 Ma) (Jensen et al., 2006; Linnemann et al., 2019), with *Parasammichnites pretzeliformis* most likely corresponding to the terminal Ediacaran (i.e. not older than 542 Ma) (Buatois et al., 2018). These trace fossils are predominantly found in micaceous fine-grained sandstone recording deposition in a shallow-subtidal environment (Buatois et al., 2018, Figure S3). *Parapsammichnites pretzeliformis* is the earliest evidence of sediment bulldozing, and documents scribbling and meandering trails with active backfill. The sporadic distribution of this trace fossil on bedding planes is interpreted to reflect the tracemakers ability to sense and exploit nutrient-rich regions (Buatois et al., 2018).

Our Cambrian specimens consist of two ichnospecies of *Psammichnites*. The older specimens, classified as *Psammichnites* cf. *saltensis*, come from member 2B (Chapel Island Formation) of Newfoundland, interpreted to include shelf to upper shoreface environments (Myrow & Hiscott, 1998; Gougeon et al., 2018; Gougeon 2023, Figure S4). Based on trace-fossil, small-shelly fossil, and acritarch data, this member is Fortunian in age (Narbonne et al., 1987). *Psammichnites* cf. *saltensis* records what has been described as a distinct, internally programmed, behavioural pattern (Seilacher, 1967; Hofmann & Patel, 1989; Hofmann, 1990).

Our youngest specimens, classified as *Psammichnites gigas circularis* (Mángano et al., 2022), are from the upper member of the Pusa Formation (Spain). This member records deposition in a shelf environment and has a Cambrian Stage 2 to Cambrian Stage 3 age (Talavera et al., 2012; Àlvaro et al., 2019; Mángano et al., 2022, Figure S5). *Psammichnites gigas circularis* has been convincingly interpreted as a searching trail of a molluscan bulldozer (Seilacher-Drexler & Seilacher, 1999; Seilacher, 2007), and commonly referred to as the "lasso trail".

Photographs of the stratigraphic surfaces, taken as close to plan view as possible, were either collected by the authors [B.A.L, R.G., M.G.M, L.A.B] (*Parapsammichnits pretzleformis, Psammichnites gigas circularis* and *Psammichnites* cf. *saltensis*) or graciously provided by Prof. Mary Droser (*Helminthoidichnites tenuis*). All specimens on the photographs were traced in Adobe Illustrator with the pen tool to produce vector images of each specimen, with trail width, scale, and direction of travel (if discernable) indicated (Supplementary Information, Figures 4.6 to 4.9). These images were subsequently imported into MATLAB and discretized according to the methodology outlined in **Chapter 3**. This methodology converts images of trace fossil paths to 2D curves that can then be subdivided into equidistant segments along the fossil trajectory according to an inferred velocity distribution.

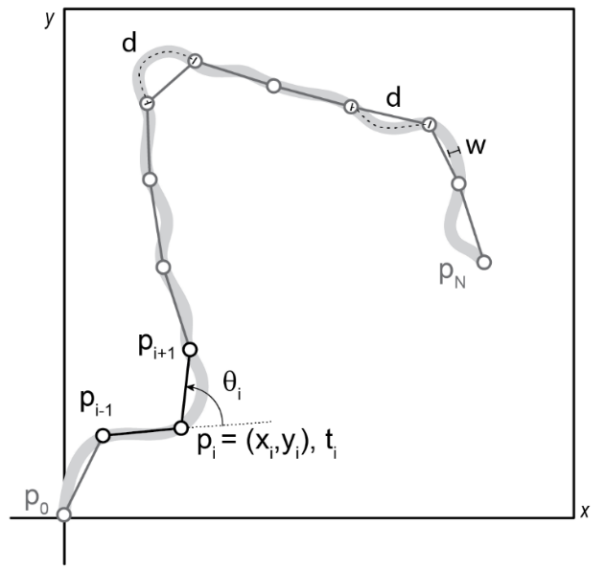

Figure 4.1. Sample movement path (thick grey line), subdivided into equidistant segments, with measures used in this study indicated: $p$ is the point data which delineates the discretized movement path (thin dark grey and black line), with associated $x$- and $y$- coordinates as well as inferred time ($t$) data, $\theta_i$ is the turning angle for the point $p_i$, $w$ is the trail width, $d$ (dotted black line) is the segment distance along the curvilinear length of the trail ($d = w*s$, where $s$ is the segment distance multiplier).

For our specimens we assumed that an average velocity provided a reasonable approximation of the velocity distributions of each tracemaker. Each equidistant segment can be interpreted to represent the passage of an approximately constant unit of time, thus providing spatial and inferred temporal data of each trajectory. From this data descriptive measures can be calculated (Jones, 1977; Marsh & Jones,1988; Calenge et al., 2009; Dray et al., 2010). For our purposes, we calculated the relative turning angle at equidistant points (*p*) along the movement path. Each point represents the location of the organism after *i* "steps". The distance between these points (*d*) is calculated by multiplying the trail width (*w*) by a segment distance multiplier (*s*) (Figure 4.1). In this way the points are spaced according to a readily available biological measurement and is resilient to fluctuations in trail width within or across populations as well as possible errors in scale (**Chapter 3**). For every specimen we collected point-data at segment distance multipliers of 0.1 and 1. Likewise, for each specimen we calculated two time-ordered series of turning angle data. The turning angle (*θ*) is the deviation in travel from the current trajectory (Figure 4.1) and is defined as the angle between three points ($p_{i-1}$, $p_i$, and $p_{i+1}$). This was calculated according to the method outlined in **Chapter 3**.

To investigate temporal trends in the fossil trajectories we sought to investigate the relationship between pairs of turning angles spaced increasingly far apart (i.e. "time lags", *h*). To do so, we calculated the autocorrelation of the turning angles of a path, where the turning angles ($\theta_i$) are compared with a lagged copy of themselves ($\theta_{i+h}$) at increasing lags (*h*). When performed on extant organism data, each lag is measured in units of time (e.g. hours) (Austin et al., 2004; Wittenmyer et al., 2008; Cushman et al, 2010; Dray et al., 2010). As we are unable to collect real time data in fossil trajectories, our lags are measured in units of distance (i.e trail width *w*). As a result, the turning angles are compared with a copy of themselves spaced *D* distance away (where *D = h·w*) for increasing values of *h*. The autocorrelation function relies on the calculation of Pearson's correlation coefficient (*r*), which produces values within the range [-1 to 1] and describes the strength and direction of the linear correlation of two variables. It is calculated by dividing the covariance of these variables by the product of their standard deviations:

$$r_{xy} = \frac{\sum_{i=1}^{N}(x_i - \bar{x})(y_i - \bar{y})}{\sqrt{\sum_{i=1}^{N}(x_i - \bar{x})^2}\sqrt{\sum_{i=1}^{N}(y_i - \bar{y})^2}} \; .$$

Where, $N$ is the sample size, $x_i$, $y_i$ are the individual sample points indexed with $i$, and $\bar{x}$ ($\frac{1}{n}\sum_{i=1}^{N} x_i$) is the sample mean (as with $y$). When applied to time-series analysis the variables $x$ and $y$ are lagged subsamples of the turning angles of a single trajectory. Therefore, for samples sizes which are large with respect to the lag analysed their means will be very similar. As a result, Pearson's correlation coefficient can be simplified to

$$r_h = \frac{\sum_{i=1}^{N-h}(x_i-\bar{x})(x_{i+h}-\bar{x})}{\sum_{i=1}^{N}(x_i-\bar{x})^2},$$

commonly known as the autocorrelation for lag $h$. Traditionally, Pearson's correlation coefficient only reflects the strength and direction of the linear relationship of the two variables. When applied to time-series analysis however, the correlation coefficient approximates the slope ($m$) of the line of best fit (least squares method):

$$m = \frac{\sum_{i=1}^{N}(x_i-\bar{x})(y_i-\bar{y})}{\sum_{i=1}^{N}(x_i-\bar{x})^2},$$

where $\bar{y} = \bar{x}$, $y_i = x_{i+h}$. As a result, the autocorrelation at lag $h$ describes the strength, direction, and slope of the linear relationship between a variable ($x_i$) and it's time-lagged counterpart ($x_{i+h}$) (Figure 4.2). An autocorrelation coefficient larger than 0 ($0 < r < 1$) indicates that the variables are correlated at lag $h$ (if $x_i$ increases it is likely that $x_{i+h}$ will increase, and and if $x_i$ decreases it is likely that $x_{i+h}$ will decrease), while a correlation coefficient smaller than 0 ($-1 < r < 0$) indicates the variables are anti-correlated (if $x_i$ increases it is likely that $x_{i+h}$ will decrease, and if $x_i$ decreases it is likely that $x_{i+h}$ will increase). Correlation coefficients near 0 indicates that the variables are uncorrelated at lag $h$. Uncorrelated, however, does not necessarily imply random as the trajectories could exhibit a non-linear non-random relationship. Additionally, Pearson's correlation coefficient is symmetric (corr(x, y) = corr(y, x), meaning a specimen will produce the same value of $r$ regardless of the direction of travel. This is particularly useful in the study of fossil trajectories as the direction of travel is often unable to be determined.

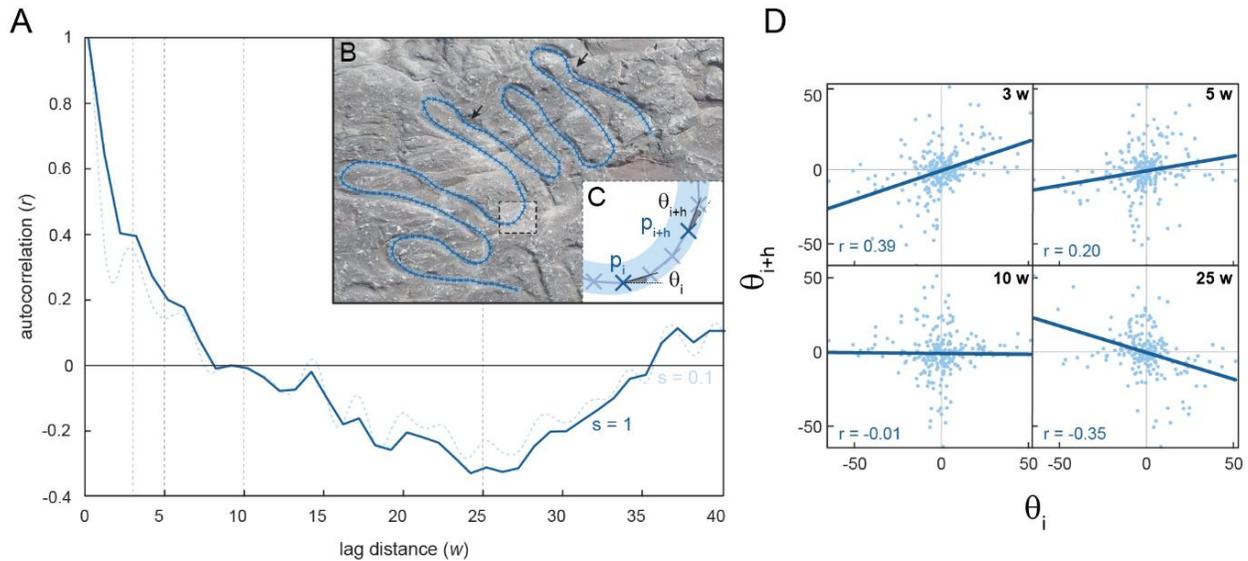

Figure 4.2. (A) Example autocorrelation function for Psammichnites cf. saltensis specimen no. 2 from member 2B of the Chapel Island Formation (early Cambrian). (B) Trajectory of P. cf. saltensis specimen no. 2, with segmented (s = 1) discretized data overlain. Dashed box shown in C. (C) Key variables used in the autocorrelation calculation (point data, p, with associated turning angles, θ). Lag distances used in B are illustrated by dotted gray lines. (B) Scatterplots of turning angles (θi) vs. their time-lagged counterparts (θi+h) for lag distances of 3, 5, 10, and 25 w (s = 1). r = autocorrelation coefficient.

To investigate the impact shorter lags were having on the autocorrelation of subsequent lags we also conducted partial autocorrelation analyses. This analysis describes the strength, direction, and slope of the linear relationship between a variable ($x_i$) and it's time-lagged counterpart ($x_{i+h}$), similar to autocorrelation, however also adjusts for the linear effects of all preceding lags ($x_{i+1}, ... x_{i+h-1}$) (Box et al., 1994).

For each individual specimen, at both segment multipliers (s = 0.1 and 1), we identified the total number of segments and calculated the autocorrelation (r) for all possible h values using MATLAB's autocorr function for autocorrelation or MATLAB's parcorr function for partial autocorrelation. This gives a series of autocorrelation values for each trajectory at each lag h within a group (i.e ichnospecies). From these values the median, 25th and 75th percentiles, and outliers were calculated and plotted as boxplots. We also calculated the mean autocorrelation (i.e. mean r) at every lag h as well as the standard deviation on the mean. The standard deviation on the mean (i.e. standard error of the mean) was calculated by dividing the standard deviation by the square root of the total number of specimens with r-values at lag h.

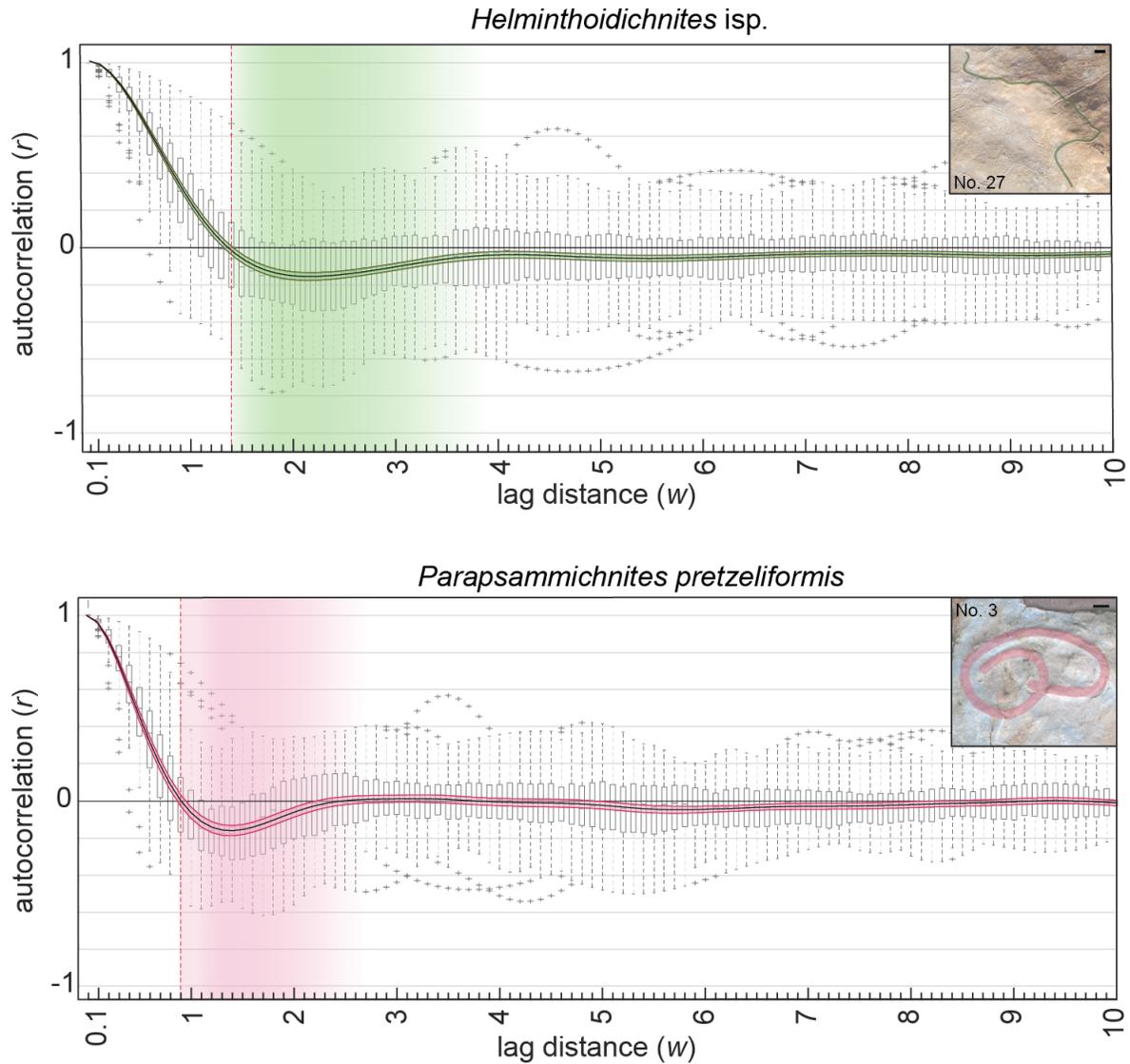

Figure 4.3. Autocorrelation functions for Ediacaran specimens, performed on turning angles collected at segment multipliers ($s$) of 0.1. Black lines denote mean $r$ values, while coloured lines indicate the standard error of the mean ($SE = \frac{\sigma}{\sqrt{n}}$). Top-right inset boxes in each graph illustrate the longest trail of each ichnospecies used in the analysis. Scales = 1 cm. Dashed red lines indicate the inferred end of the sampling-induced positive autocorrelation. Shaded intervals indicated regions of interest discussed in the text. Mean $r$ and SE data available in <mark>Supplementary Data</mark>. for *Helminthoidichnites tenuis* and *Parapsammichnites pretzeliformis* respectively.

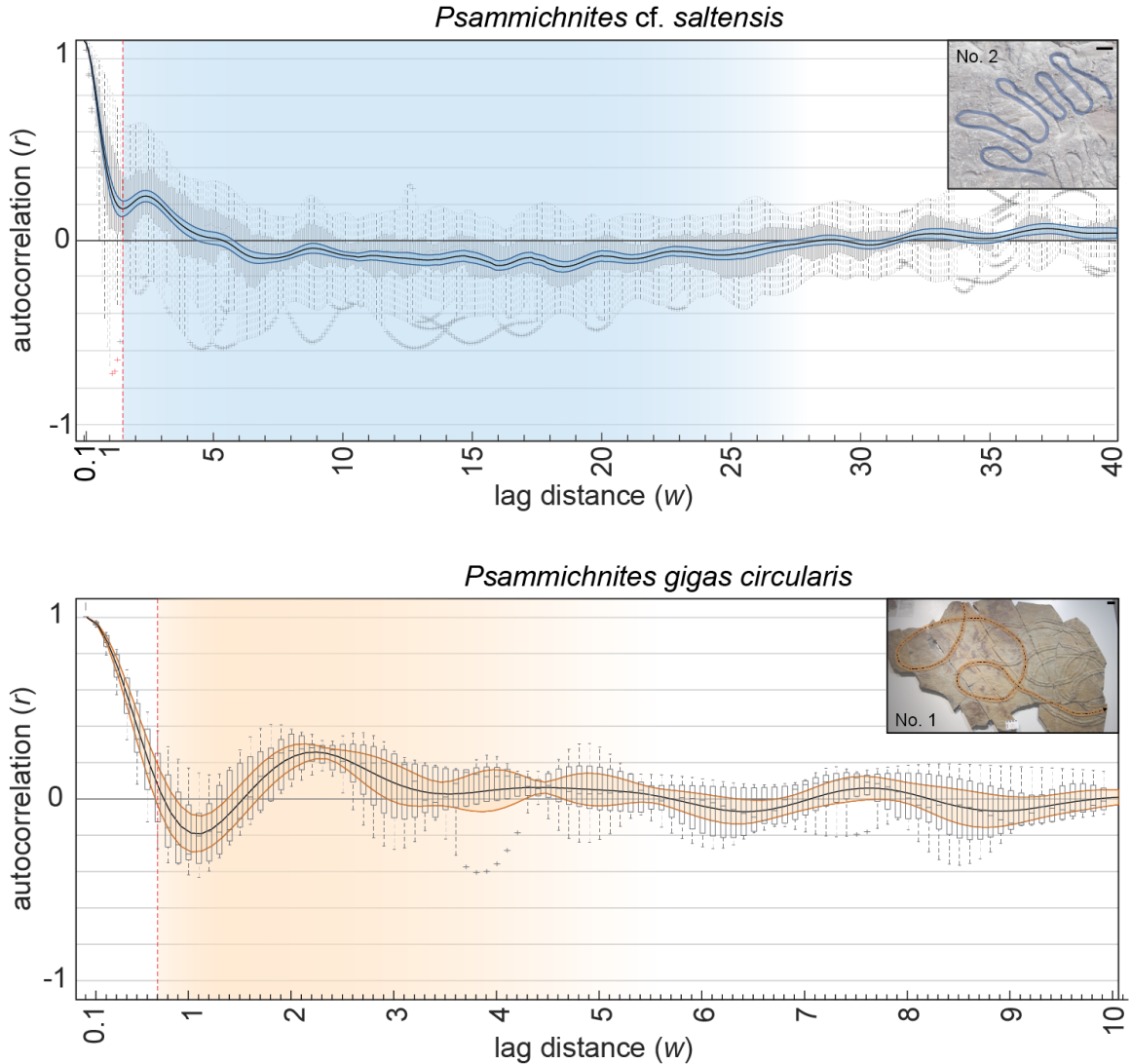

Figure 4.4. Autocorrelation functions for Cambrian specimens, performed on turning angles collected at segment multipliers ($s$) of 0.1. Black lines denote mean $r$ values, while coloured lines indicate the standard error of the mean ($SE = \frac{\sigma}{\sqrt{n}}$). Top-right inset boxes in each graph illustrate the longest trail of each ichnospecies used in the analysis. Scale for *Psammichnites* cf. *saltensis* = 1 cm, scale for *Psammichnites gigas circularis* = 5 cm. Dashed red lines indicate the inferred end of the sampling-induced positive autocorrelation. Shaded intervals indicated regions of interest discussed in the text. Note the difference in x-axis scale for *Psammichnites* cf. *saltensis*. Mean $r$ and SE data available in Supplementary Data for *Psammichnites* cf. *saltensis* and *Psammichnites gigas circularis* respectively.

4.4 RESULTS

Each autocorrelation plot shows a decreasing positive correlation at small lags (0.8 *w* to 1.5 *w*) (Figure 3.4 & 4.4). This is an artefact of the short segment distance multiplier used (*s* = 0.1). These lag distances are likely shorter than the distance covered by the organism in a single action or swimming stroke, resulting in autocorrelation that does not reflect the navigational choices of the tracemaker. Similar issues arise with the use of GPS data on extant organisms, where decreasing time intervals see an increase in the dependence between observations (Dray et al., 2010). The distance this autocorrelation ends at, therefore, could indicate the typical distance travelled in a single action by the organism. Turning angles can be subsequently re-calculated and re-analysed by these "action steps" (Figure 4.5) and demonstrate a similar trend to that observed in the lag distance plots (Figure 4.3 & 4.4).

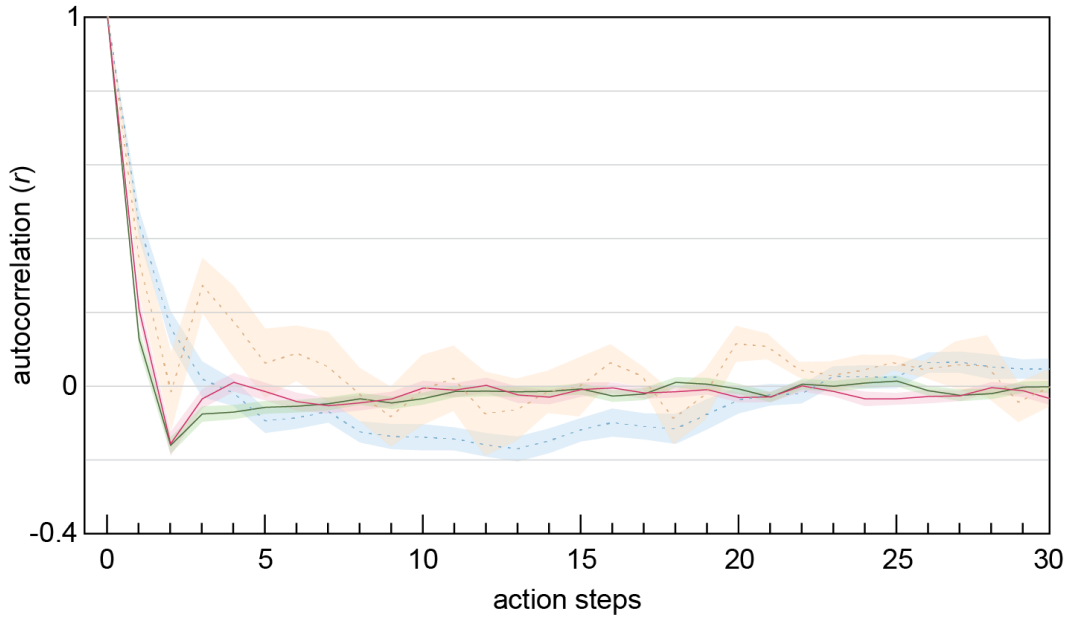

Figure 4.5. Autocorrelation functions. Solid lines = mean *r*, shaded coloured regions indicate the standard error of the mean ($SE = \frac{\sigma}{\sqrt{n}}$). X-axis is the "action step", determined by the sampling-induced autocorrelation distance (red dashed lines in Figures 4.3 & 4.4). Green is *Helminthoidichnites tenuis* (action step = 1.4 *w*), Pink is *Parapsammichnites pretzeliformis* (action step = 0.9 *w*), dashed blue is *Psammichnites* cf. *saltensis* (action step = 1.4 *w*) and dashed orange is *P. gigas circularis* (action step = 0.7 *w*).

Immediately after the inferred sampling-induced autocorrelation the mean *r* values of *Helminthoidichnites tenuis*, and *Parapsammichnites pretzeliformis* become weakly negative, indicating possible anticorrelation in the lagged turning angles (Figure 4.3). In *P. pretzeliformis*, 75% of paths have negative correlation coefficients between lag distances of 1.1 *w* and 1.5 *w*, with a minimum mean *r* of -0.16 (lag distance = 1.4 *w*). The mean *r* for *P. preztlformis* remains negative from a time lag of 1 *w* until around 2.4 *w*. In *Helminthoidichnites tenuis*, positive mean *r* values persist from lag 1.4 *w* to about 3.4 *w*, although with greater variability than in *P. pretzeliformis*, with a minimum mean *r* of -0.15 (lag distance = 2.0 *w*). The minimum mean r for both ichnospecies is below the generally accepted threshold for weak anticorrelation (0.2 < *r* < 0.4), however these intervals are still distinct from the remainder of the autocorrelation function in *Helminthoidichnites tenuis* and *P. pretzeliformis*. In turn, the standard deviation on the mean for both ichnospecies remains small, providing greater evidence in favour of significant anticorrelation at these time lags. After these brief intervals of tentative anticorrelation the mean autocorrelation for both ichnospecies remains near 0. The lagged turning angles therefore are, on average, uncorrelated past this interval. This indicates that turns spaced greater than 3.4 or 2.4 trail widths apart are unlikely to be correlated for *Helminthoidichnites tenuis* and *P. pretzeliformis*, respectively. In terms of inferred action-steps, both *Helminthoidichnites tenuis* and *P. pretzeliformis* trajectories are tentatively anticorrelated after 2 action steps and show no demonstrable autocorrelation or partial autocorrelation past 3 action steps (Figures 4.5 & 4.6, and Supplementary Information, Figures 4.10 & 4.11).

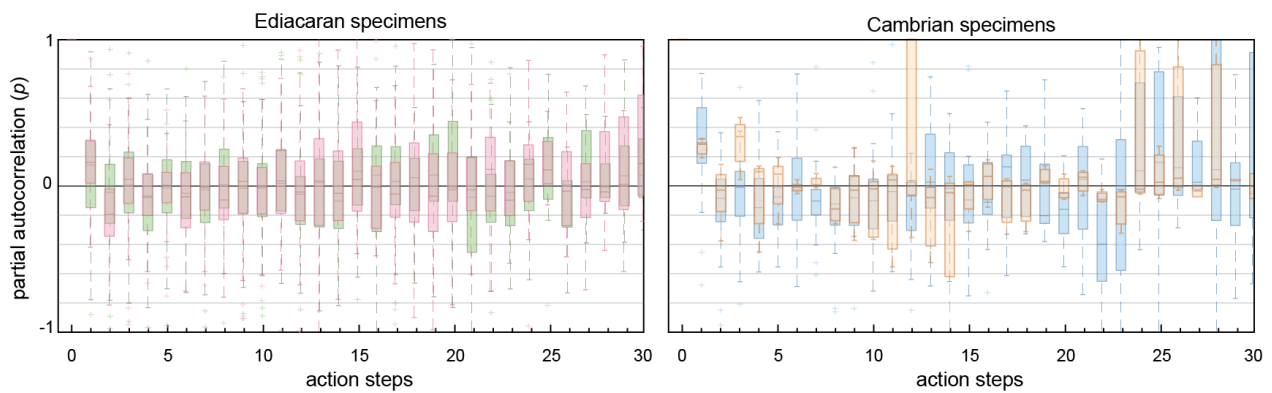

Figure 4.6. Partial autocorrelation boxplots for the sampled Ediacaran specimens and Cambrian specimens. X-axis is the "action step", determined by the sampling-induced autocorrelation distance (red dashed lines in Figures 4.3 & 4.4). Green is *Helminthoidichnites tenuis* (action step = 1.4 w), Pink is *Parapsammichnites pretzeliformis* (action step = 0.9 w), dashed blue is *Psammichnites* cf. *saltensis* (action step = 1.4 w) and dashed orange is *P. gigas circularis* (action step = 0.7 w).

In contrast, the mean *r* values of early Cambrian specimens of *P*. cf. *saltensis* becomes increasingly positive after the sampling-induced autocorrelation (Figures 4.2 & 4.4). At least 75% of the paths demonstrate positive autocorrelation until a lag of 3.4 *w*, with a maximum mean *r* of 0.22 (lag distance = 2.4 *w*). The mean correlation coefficient remains positive until a lag of 5.4 $w$ or 2 action steps. Similarly, at least 75% of paths demonstrate positive partial autocorrelation until a lag of 1 action step (Figure 4.6 and Supplementary Information, Figures 4.10 & 4.11). This is immediately followed by a long interval of negative mean $r$ values which fluctuate around a mean $r$ of -0.1 (min of -0.13, lag distance = 18.5 $w$) until they approach 0 around a lag of 28 $w$. For most of this interval, around 75% of specimens have a negative correlation coefficient, although partial autocorrelation values fluctuate around 0 until 22 actions steps, where at least 75% of specimens have a negative partial autocorrelation (Figure 4.6 and Supplementary Information, Figures 4.10 & 4.11). This indicates that, at the studied locality, the *P.* cf. *saltensis* tracemaker was likely to follow a similar turning angle after travelling short distances (up to 5.4 $w$, or 2 action steps), and then had a preference to make an opposing turn after travelling between 5.4 $w$ to 28 $w$ (or 4 to 22 action steps). Notably, partial autocorrelation values show demonstrable variance after 22 action steps

Like *Helminthoidichnites tenuis* and *Parapsammichnites pretzeliformis*, the mean autocorrelation of *Psammichnites gigas circularis* also becomes negative shortly after the sampling-inducing autocorrelation (Figure 4.4). At least 75% of the paths have negative autocorrelation for lag distances of 1.0 $w$ to 1.2 $w$, with a minimum mean $r$ of -0.19 (lag distance = 1.1 $w$). The mean autocorrelation remains negative until a lag distance of 1.6 $w$. This is immediately followed by an interval of positive mean $r$ values, which continue to increase until a maximum mean $r$ of 0.26 (lag distance = 2.2 $w$) and then decrease until a mean $r$ of 0.03 (lag distance = 3.3 $w$). Between lag distances of 1.9 $w$ and 2.5 $w$, 100 % of *P. gigas circularis* specimenswere positively correlated, with mean $r$ values between 0.20 and 0.26. Similarly, 100% of specimens showed positive partial autocorrelation after 3 action steps (2.1 $w$). Each turn, therefore, is weakly correlated ($0.2 < r < 0.4$) at these time lags. This indicates that the *P. gigas circularis* tracemaker was likely to make an opposite turn after travelling 1 to 2 $w$ (2 action steps), followed by a similar turn after travelling 2 to 3 $w$ (3 action steps). Notably, partial autocorrelation values for *Psammichnites* cf. *saltensis* and *Psammichnites gigas circularis* specimens show demonstrable variance after 22 action steps (and at 12 action steps for *P. gigas circularis*), a trend that was not seen in the studied Ediacaran specimens (Figure 4.6 and Supplementary Information, Figures 4.11 & 4.12).

## 4.5 DISCUSSION

We predicted that our Ediacaran tracemakers would not utilize time-tuned behaviours in the creation of their trajectories, but that our Cambrian tracemakers would. Our results seem to uphold this initial conjecture, with a lack of significant temporal autocorrelation in the movement trajectories of the two Ediacaran ichnotaxa analysed after short distances, but distinct autocorrelation in the analysed Cambrian ichnotaxa.

Whilst this does not necessarily mean that the Ediacaran tracemakers lacked navigatory systems capable of generating time-tuned behaviours, it does imply that this information did not affect their subsequent trajectories. The tentative anti-correlation in the Ediacaran autocorrelation functions at short lag distances may suggest a course-correcting strategy or reflect anatomical impacts on motion (i.e., the gait of a single swimming stroke) of the *Helminthoidichnites tenuis* and *P. pretzeliformis* tracemakers, applied over short distances (max lag distance = 2.5 to 3.5 *w*). However, the results for the Cambrian tracemakers do suggest past movements are correlated to future movements for short (*P. gigas circularis*, max lag distance = 3 *w*) or moderate distances (P. cf. *saltensis*, max lag distance = 30 *w*). This indicates either the evolution of time-tuned behaviours for these Cambrian grazers, or, most likely, an increased necessity to apply them. This increased necessity could be due to different environmental conditions, from shallow marine and subtidal environments for our Ediacaran specimens to shelf and upper shoreface environments for our Cambrian ones, or due to the increased ecological complexity and novel ecosystems thought to appear during the Ediacaran-Cambrian transition.

One adaptation that could result in the time-tuned behaviours observed in our Cambrian tracemakers is the evolution of biologic memory. The metabolic cost for memory systems can be low. External spatial memory systems have been observed in nonneuronal organisms (e.g. the slime mold *Physarum polycephalum*), in which the secretion and detection of chemical markers (e.g. extracellular slime) assists the organism to avoid previously-explored areas (Reid et al., 2012).When simulated chemical markers were allowed to decay, simulated foragers were able to utilize an external memory system that informed them not only if they had visited a site before (spatial memory) but when (temporal memory) (Chung & Choe, 2009). This external system of memory requires little metabolic overhead, utilizes an ancient sensory modality (i.e. chemoreception), and is thought to be a precursor to more complex internalized systems of memory (Hildebrand, 1995; Chung & Choe, 2009; Reid et al., 2012). Koy and Plotnick (2007) simulated a similar external memory system, though without the decay component, by programing a "tracemaker" to first sense chemical gradients and then to respond to them by moving towards the gradient and ingesting what was present. This program causes previously travelled regions to have decreased nutrient concentrations, effectively serving as a chemical marker in an external memory system. They demonstrated how this single strategy can produce different trajectories depending on the distribution of chemical gradients. In other words, the final trajectories differed due to the external environment, not due to differences in the navigation or motion capacities of the tracemaker. A similar system could account for the appearance of avoidance behaviours in Ediacaran and Cambrian grazers such as *Helminthoidichnites* and *Psammichnites*, including occasional avoidance behaviours in our trajectories of *P.* cf. *saltensis* (Figure 4.2, Carbone & Narbonne, 2014; Gehling & Droser, 2018).

Such external systems of memory respond on an action-by-action basis to the external environment. While changes in external conditions will have an impact on the final trajectory, it is unlikely that external factors such as nutrient distribution, navigated via an external memory system, can solely account for the autocorrelation seen in our Cambrian trajectories. For this system to produce autocorrelated trajectories the external environment would need to be highly structured. This may be possible in situations where grazing behaviour is dependent on highly periodic external conditions (e.g. circadian or circatidal clock-driven behaviours), though we contend that such influences would generate impacts at a larger scale than the high-resolution motion examined here (i.e. turning angles). As a result, we interpret the presence of autocorrelation in our Cambrian trajectories indicated the tracemaker followed a behavioural strategy that was dependent on a past feature, be it past actions, environmental cues, or a change in internal state. This suggests a time-tuned behaviour, where the organism followed a specific trajectory or rhythmic behaviour for (or after) a set amount of time after the detection of a cue. For meandering fossil trajectories, like *P.* cf. *saltensis*, the length of the tracemaker has been suggested to control the time between U-turns, with the tail-straightening serving as a cue for the formation of the next U-turn (Seilacher, 1967). While we cannot exclude this mechanism as a possibility, this does not explain the autocorrelation and partial autocorrelation seen in *P. gigas circularis*, nor the variance in the straight-line length between U-turns in *P.* cf. *saltensis*. Other methods  which produce time-tuned behavioursare more likely. In some algae, the triggering of a biological timer mediates the duration or lag-time of a response (e.g. rate of flagellar beating in the phototactic response of *Chlamydomonas*; Leptos et al., 2023). Other organisms possess clock genes which reference external time-cues (e.g. light) to help tune specific behaviours to environmental cycles (e.g. circadian, circatidal) (Naylor, 1988; Dunlap, 1999; Lakin-Thomas & Brody, 2004).

The pervasiveness of time-keeping systems throughout the tree of life demonstrate that simple time-tuned behaviours do not necessitate complex neurologic structures or somatosensory systems (Dunlap et al., 1999). Such behaviours in benthic foragers, then, could have evolved rapidly in response to changing environments. The effectiveness of biologic memory on foraging, which could produce time-tuned behaviours, is mediated by resource distribution and predictability. The use of an external spatial memory system increases the effectiveness of navigation of the slime mold *Physarum polycephalum* in complex environments (U-shaped traps) but has little impact in simple environments (Reid et al., 2012). Similarly, in homogeneous or highly unpredictable environments, biologic memory would offer little advantage. As the spatiotemporal complexity of an environment increases, the functional utility of memory also increases (McNamara & Houston, 1987; Fagan et al., 2013). As a result, animals with more complex cognitive systems do not always utilize memory-informed searches but instead switch to memory-less strategies in specific environmental conditions or internal states such as random walks (Zollner & Lima, 1999; Bartumeus et al., 2005; Codling et al., 2008).

All four of the studied trajectories may have been constructed by bilaterian tracemakers. Specifically, recent research has associated trajectories of *Helminthoidichnites tenuis* from the Ediacaran Member at Nilpena in Southern Australia with the putative early bilaterian *Ikaria wariootta* (Evans et al., 2020) and trajectories of *P. pretzeliformis* from the Spitskop Member (Urusis Formation) of Namibia with bilaterian sediment bulldozers (Buatois et al., 2018). It is probable then, that these tracemakers possessed neurons and neuromodulators capable of triggering and producing time-tuned behaviours and possibly even a biologic memory system (Sawin et al., 2000; Denes et al., 2007; Day & Sweatt, 2010; Bennet, 2021). Our results suggest, however, that the utilization of such adaptations did not offer an advantage when grazing in the sampled Ediacaran environments. It follows that our autocorrelated Cambrian trajectories can be understood as an adaptive response to spatiotemporally complex environments. This aligns with the hypothesis that increased spatial complexity in the Cambrian drove the evolution of sensory systems (Plotnick et al., 2010; Carbone & Narbonne, 2014).

Previous work on the evolutionary trajectory of horizontal pascichnial paths has relied on the qualitative description of their morphology (e.g. irregular, circular scribbles, looping, first or second order meanders) (Seilacher, 1967; 1977; 1997; Kitchell, 1979; Crimes, 1992; Crimes & Droser, 1992). Interestingly, both Ediacaran ichnospecies have extremely similar autocorrelation and partial autocorrelation functions, despite differences in trace-fossil morphology, size, mode of sediment interaction, and location (Figure 4.3 & 4.6). Meanwhile, *Parapsammichnites pretzeliformis* and *Psammichnites gigas circularis*, which have a similar "looping" morphology and mode of sediment interaction, have quite different autocorrelation and partial autocorrelation functions (Figure 4.3,4.4 & 4.6). It is apparent, then, that at least some classic descriptors of pascichnial trace-fossil morphology are not necessarily associated with the presence of time-tuned behaviours or possible memory systems.

## 4.6 FUTURE PROSPECTS

Further studies are needed to extricate the evolutionary and environmental controls on the presence of time-tuned behaviours in the Ediacaran-Cambrian transition. An extensive examination of temporal trends in Ediacaran and Cambrian movement trajectories across varying depositional environments could help constrain the timing and ecological context of the manifestation and pervasive use of time-tuned behaviours and possible biological memory. Another interesting approach would be to analyse movement trajectories for periodicity, which would allow for further examination of the relationship between internal clocks and rhythmic behaviour as well as the presence of different gaits of swimming strokes. Where large bedding surfaces are available, the spatiotemporal distribution of time-tuned trajectories (of the same ichnotaxa) could provide insight on when organisms switched between different behaviours (e.g. intra- and inter- patch strategies). Time-tuned behaviours and possible memory-based strategies are one facet of navigational complexity. Investigations on stochastic search strategies (i.e. random walks) in Ediacaran-Cambrian transition movement trajectories could reveal additional ways of increasing foraging efficiency in heterogeneous environments without the need for time-keeping systems or biologic memory.

## 4.7 CONCLUSION

The analyses presented herein have demonstrated the presence of temporal autocorrelation in the movement trajectories of two Cambrian ichnotaxa, *Psammichnites gigas circularis* and *Psammichnites* cf. *saltensis*. Trajectories of *P.* cf. *saltensis* from the Fortunian of Newfoundland showed anticorrelation up to a distance of 28 times the path width. Our analysis of two Ediacaran ichnotaxa *Helminthoidichnites tenuis* and *Parapsammichnites pretzeliformis* revealed no significant autocorrelation or partial autocorrelation at short distances but may suggest an initial course-correcting strategy. In turn, our work shows that the length of a tracemaker's "action step" (i.e. the distance travelled in a single swimming stroke) may be inferred by the length of sampling-induced positive correlation in the autocorrelation function. Finally, our results show that superficially similar trajectories (i.e. looping) can have different temporal patterns, revealing greater variation in the navigational capacity reflected by superficially similar trajectories.

Taken together, our results suggest that the appearance of time-tuned behaviours and possible memory-driven grazing strategies was not concomitant with the appearance of bilaterians but was instead an adaptive response to increasingly complex information landscapes in the early Cambrian as predicted by the Cambrian Information Revolution hypothesis. Our data supports the view that the appearance and increase in cognitive complexity and behavioural adaptations was an important evolutionary event during the Ediacaran-Cambrian transition and charts a course for the use of quantitative fossil trajectory analysis to further calibrate the timeline of sensory and nervous system innovations.

## 4.8 ACKNOWLEDGMENTS

We would like to thank M. Droser for supplying photographs of *Helminthoidichnites tenuis* specimens. B.A.L is supported via a NSERC PGS-D grant and a Macquarie iMQRES scholarship. M.G.M and L.A.B thank financial support provided by Natural Sciences and Engineering Research Council (NSERC) Discovery Grants 311727-15/20, and 311726–13 and 422931-20, respectively. M.G.M also thanks funding by the George J. McLeod Enhancement Chair in Geology.

## 4.10 SUPPLEMENTARY INFORMATION

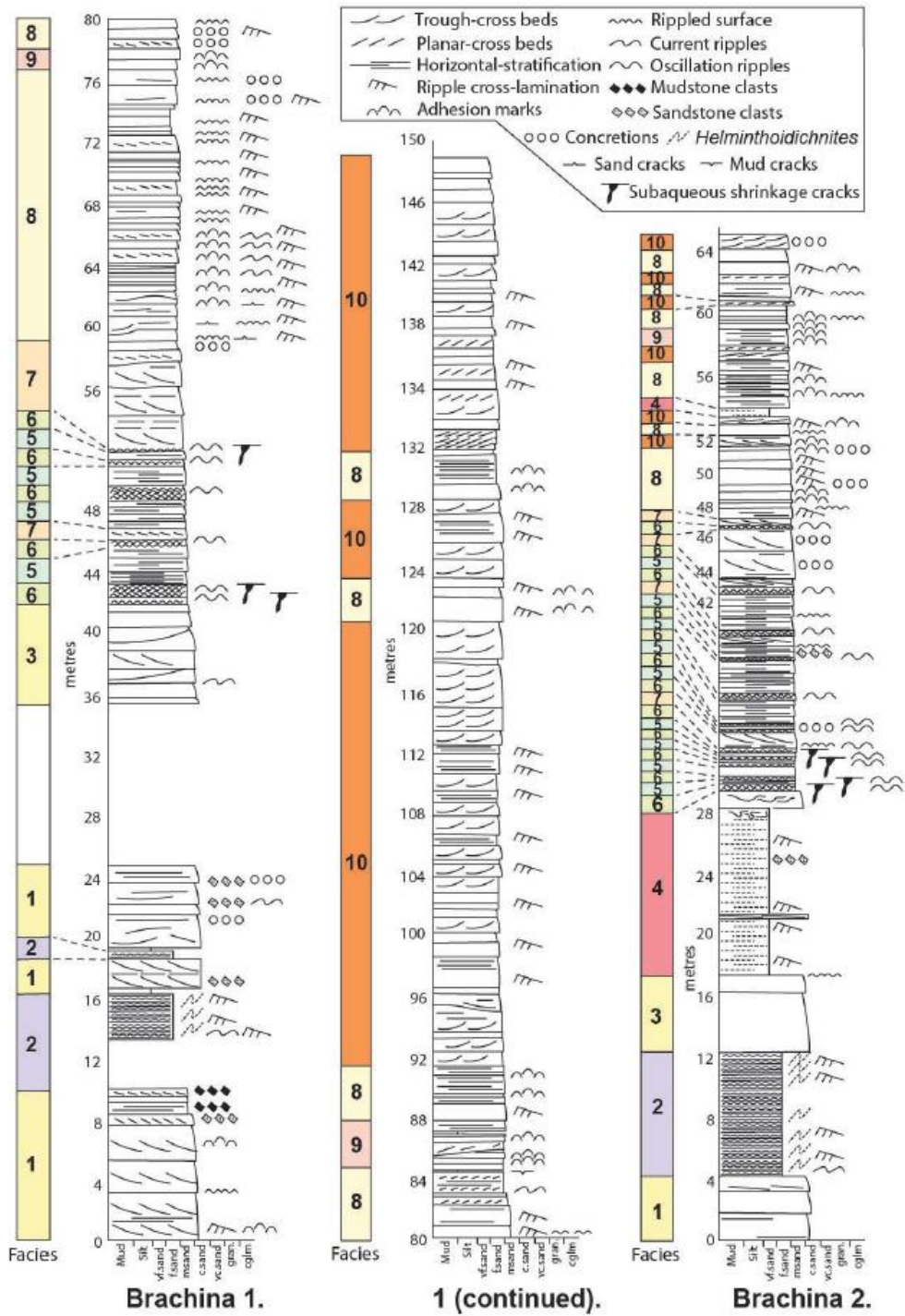

Figure 4.7. Stratigraphic logs of the Rawnsley Quartzite (Ediacara Member and upper Rawnsley Quartzite). From McMahon et al. 2020.

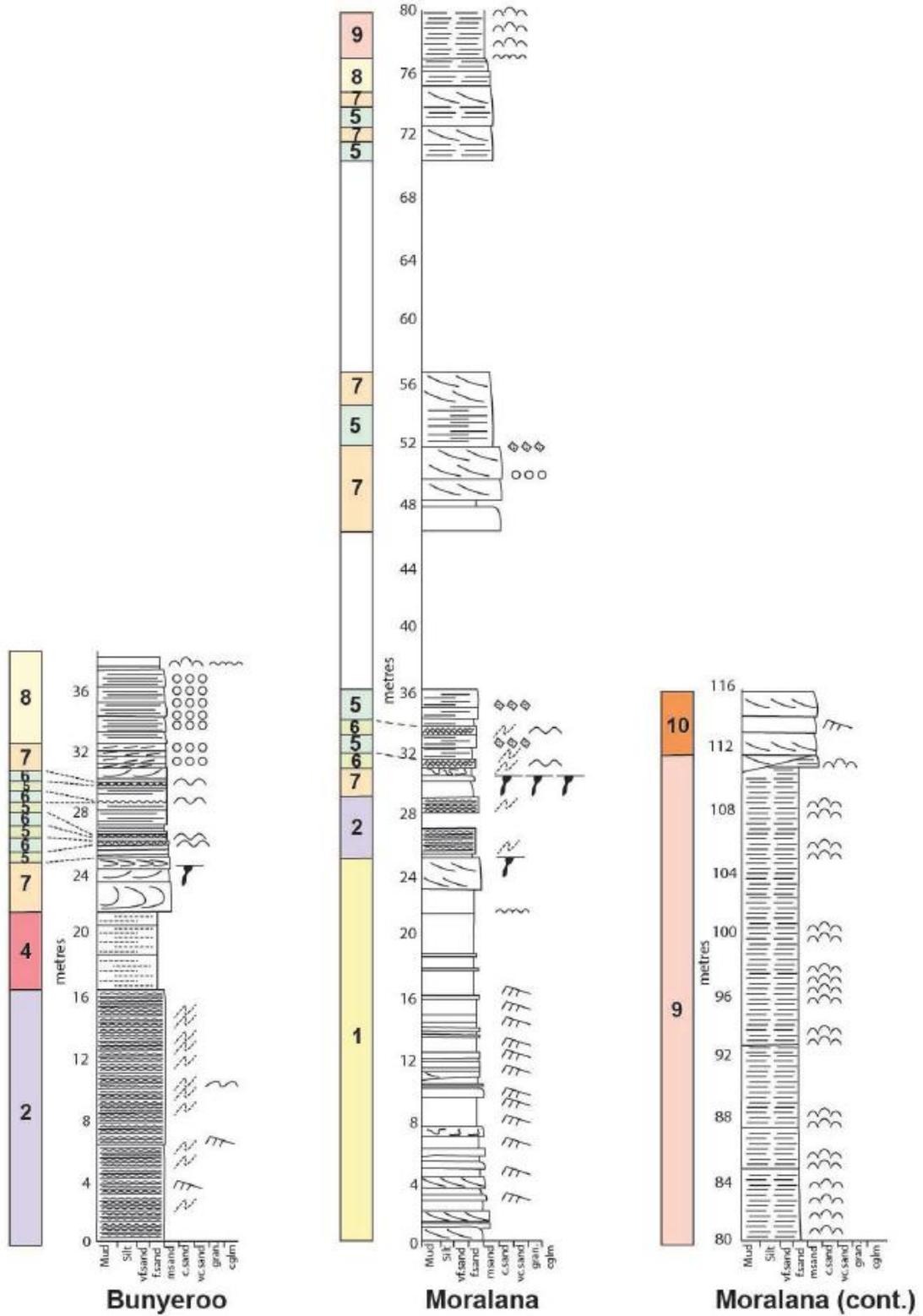

Figure 4.8. Stratigraphic logs of the Rawnsley Quartzite (Ediacara Member and upper Rawnsley Quartzite). From McMahon et al. 2020.

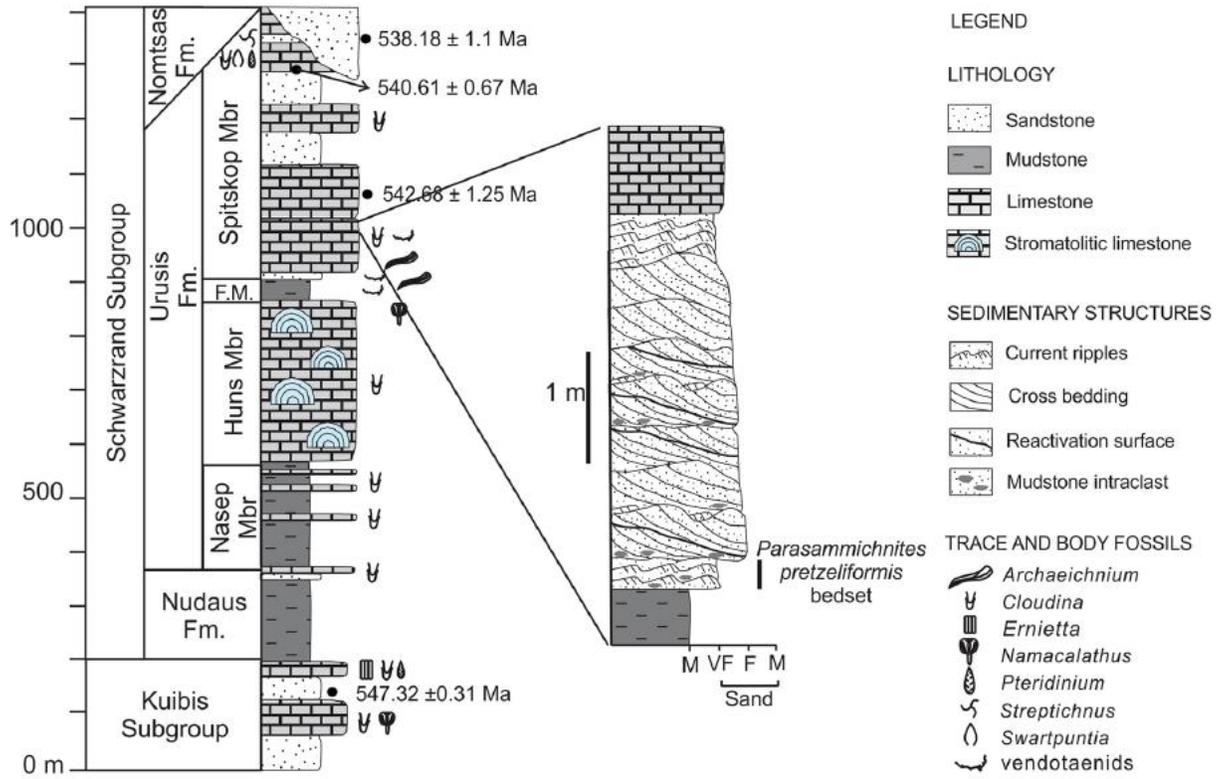

Figure 4.9. Stratigraphic section of the Schwarzrand Subgroup

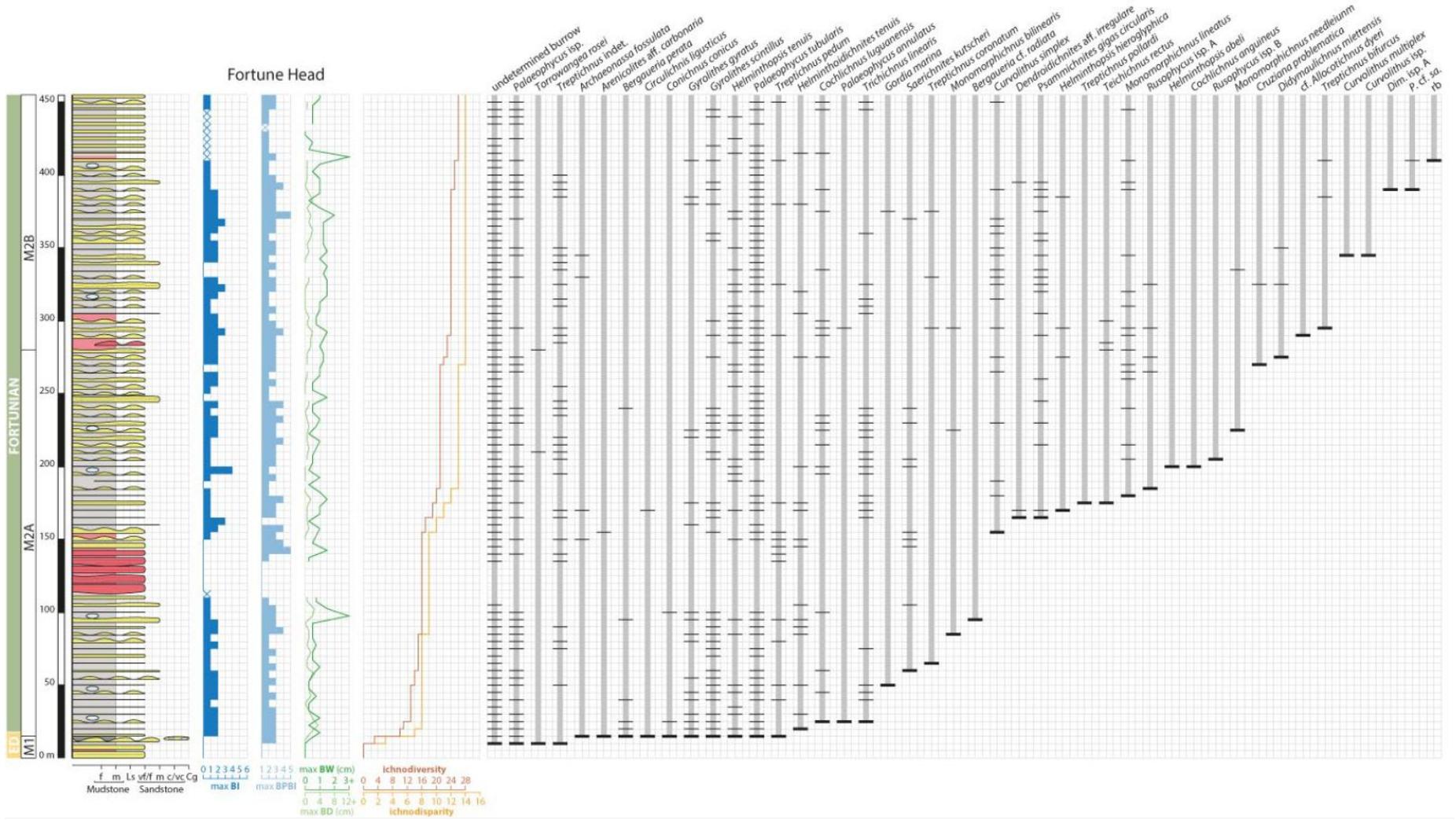

Figure 4.10 Stratigraphic log of the Chapel Island Formation from the Fortune Head locality, with interval containing the studied *Psammichnites* cf. *saltensis* speciemens indicated. From Gougeon 2023.

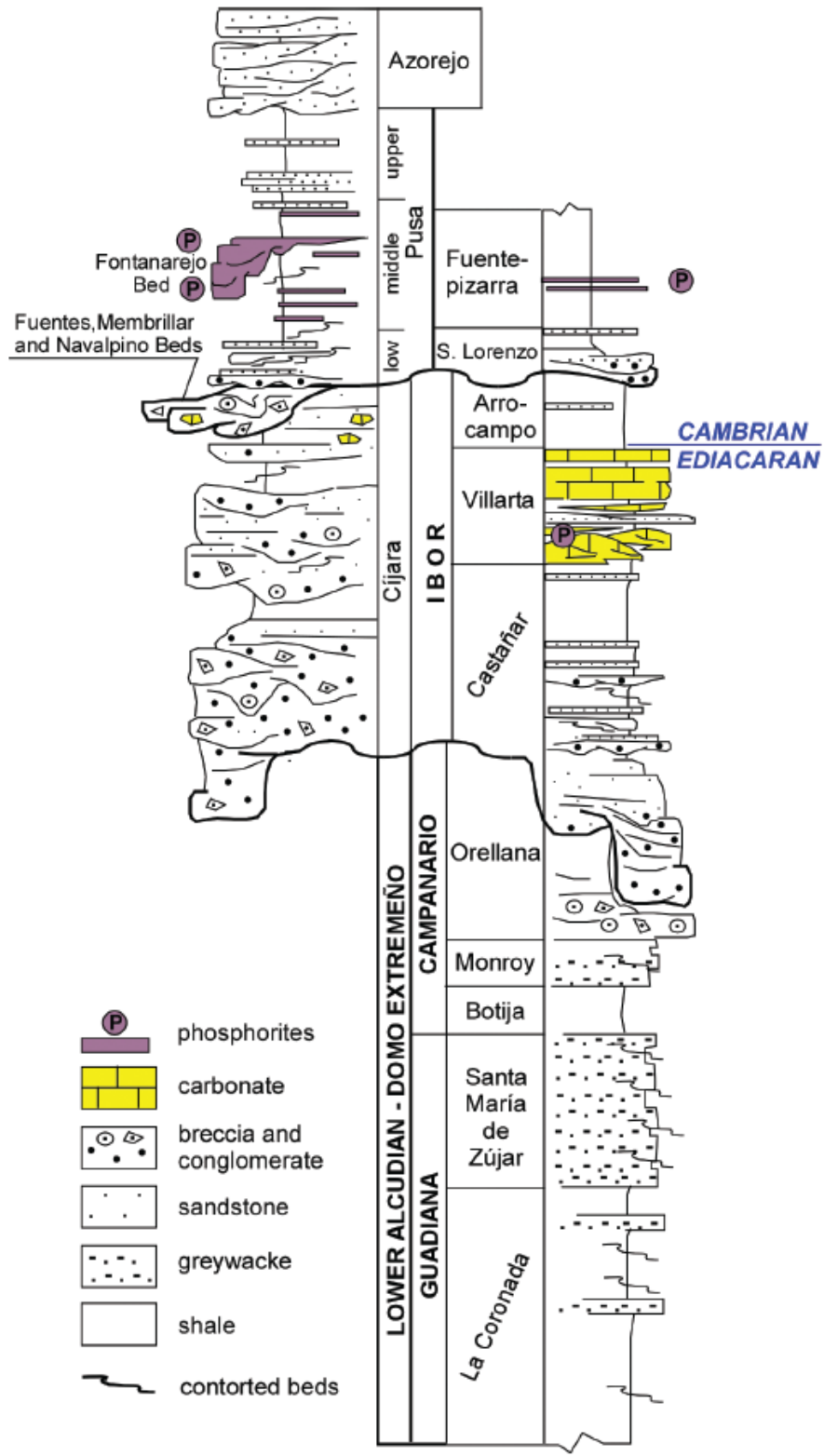

Figure 4.11. Stratigraphic log of Ediacaran and Cambrian strata in Spain. From Àlvaro et al. 2019.

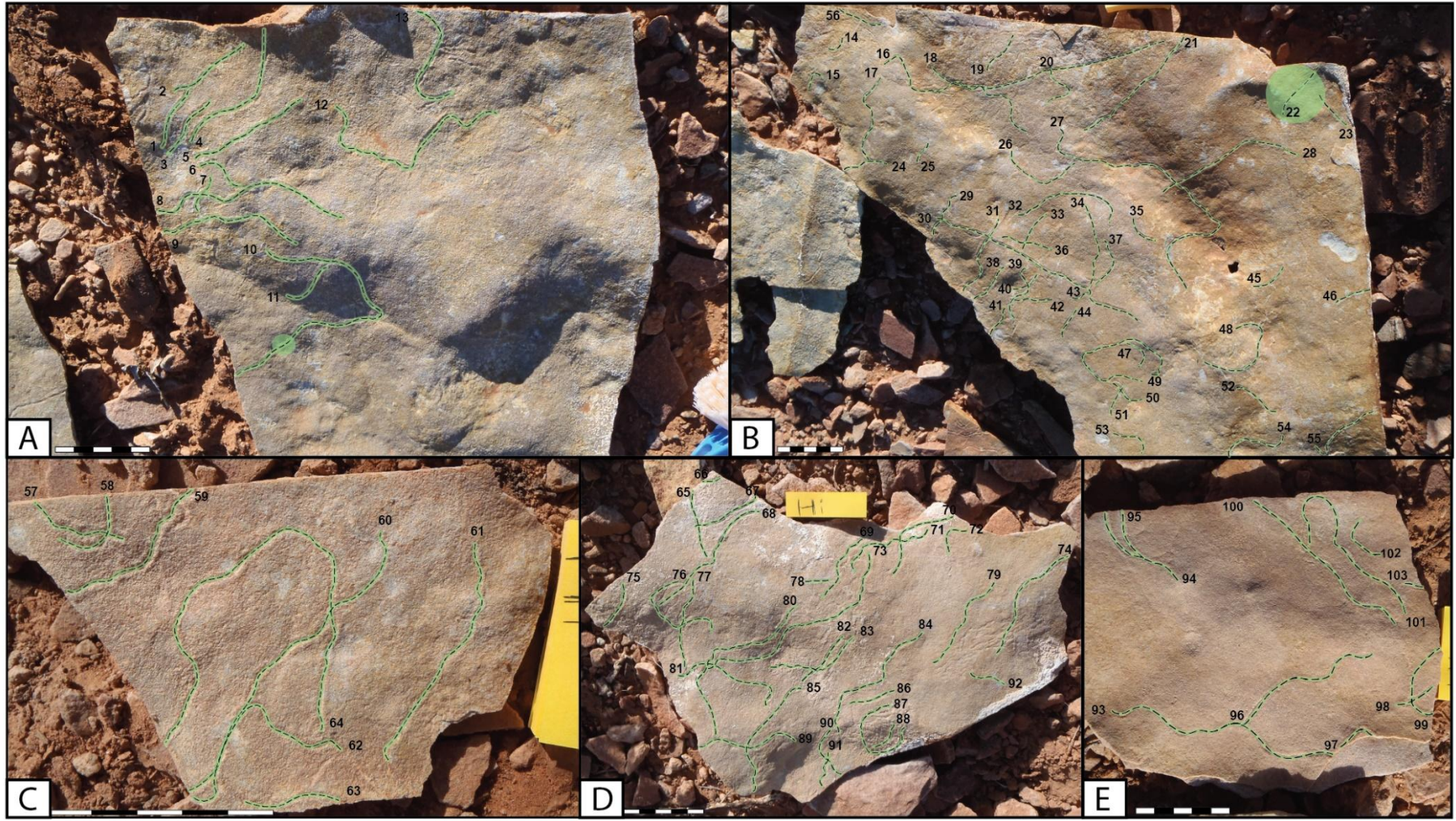

Figure 4.12. All specimens of *Helminthoidichnites tenuis*, all scale bars are 5 cm. Movement trajectories are denoted in dashed black lines, translucent green lines indicate trail width, while translucent green circles indicates specimens of *Dickinsonia*. Photos property of Dr. Mary Droser.

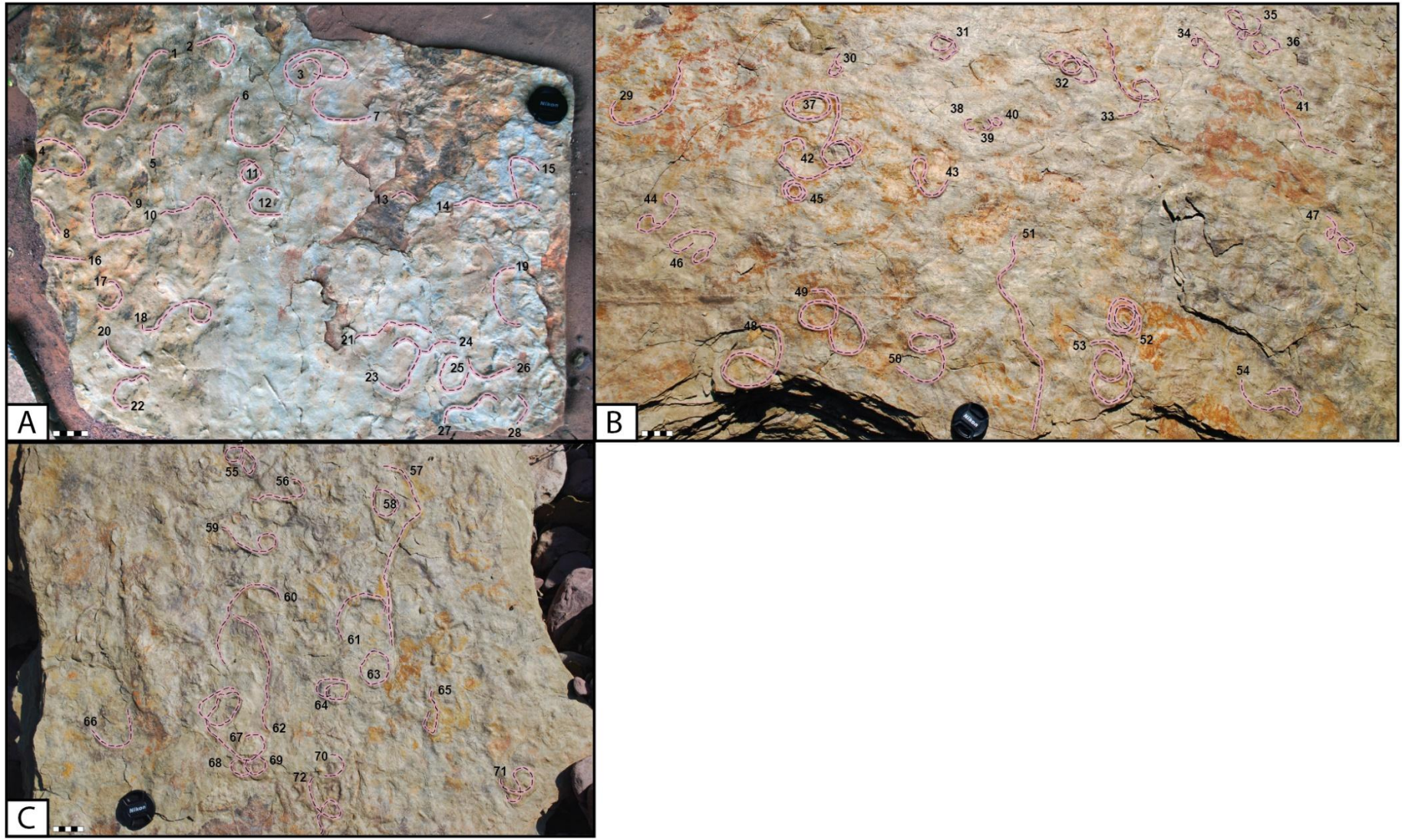

Figure 4.13. All specimens of *Parapsammichnites pretzeliformis,* all scale bars are 5 cm. Movement trajectories are denoted in dashed black lines, translucent pink lines indicate trail width.

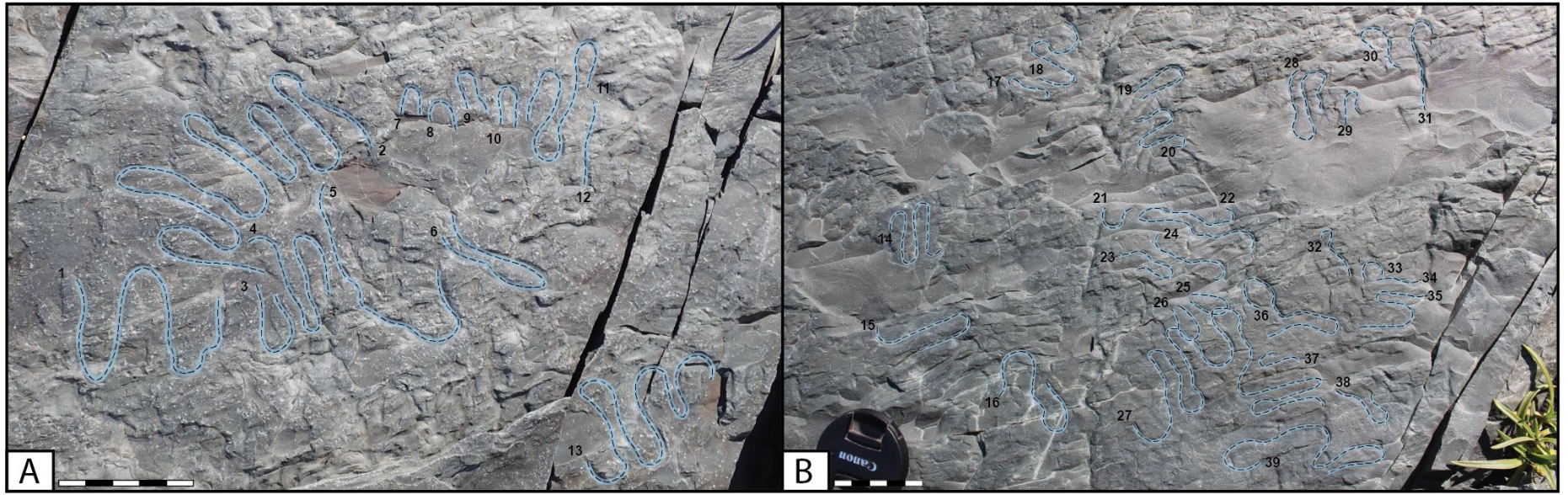

Figure 4.14. All specimens of *Psammichnites* cf. *saltensis*, all scale bars are 5 cm. Movement trajectories are denoted in dashed black lines, translucent blue lines indicate trail width.

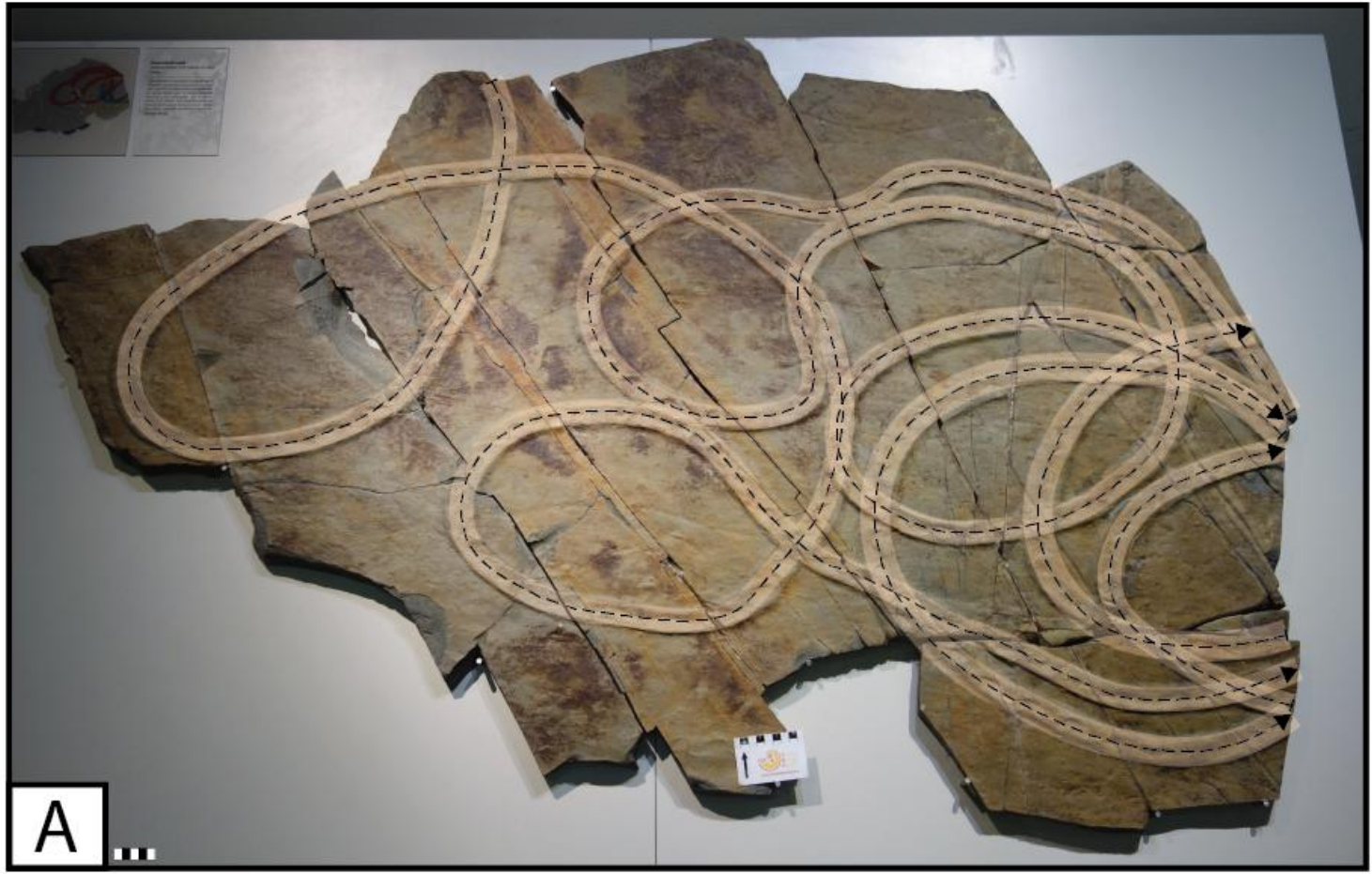

Figure 4.15. All specimens of *Psammichnites gigas gigas,* scale bar is 5 cm. Movement trajectories are denoted by dashed black lines, arrow indicates travel direction. Translucent orange lines indicate trail width

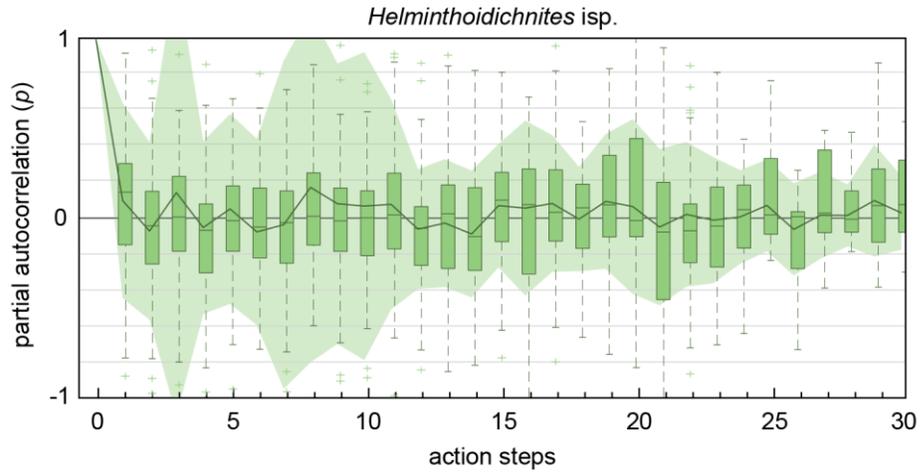

Figure 4.16. Partial autocorrelation functions for *Helminthoidichnites tenuis*, performed on turning angles collected at "action step" ($s = 1.4\ w$), determined by the sampling-induced autocorrelation distance (red dashed line in Fig. 3). Black lines denote mean $p$ values, while translucent fill indicates the standard error of the mean ($SE = \frac{\sigma}{\sqrt{n}}$).

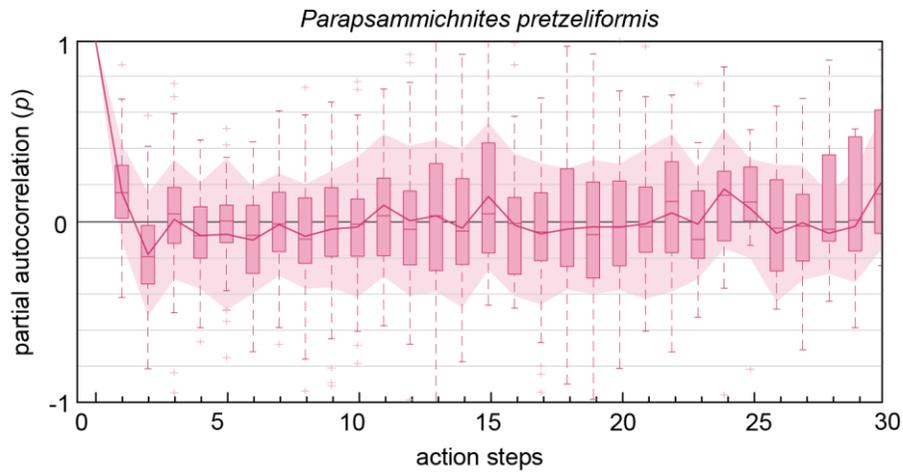

Figure 4.17. Partial autocorrelation functions for *Parapsammichnites pretzeliformis*, performed on turning angles collected at "action step" ($s = 0.9\ w$), determined by the sampling-induced autocorrelation distance (red dashed line in Fig. 3). Black lines denote mean $p$ values, while translucent fill indicates the standard error of the mean ($SE = \frac{\sigma}{\sqrt{n}}$).

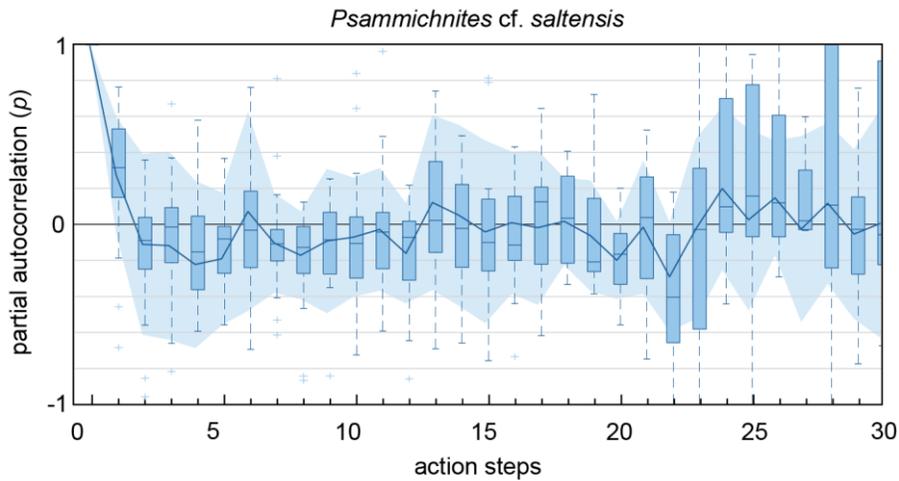

Figure 4.18. Partial autocorrelation functions for *Psammichnites* cf. *saltensis*, performed on turning angles collected at "action step" ($s = 1.4\ w$), determined by the sampling-induced autocorrelation distance (red dashed line in Fig. 4). Black lines denote mean $p$ values, while translucent fill indicates the standard error of the mean ($SE = \frac{\sigma}{\sqrt{n}}$).

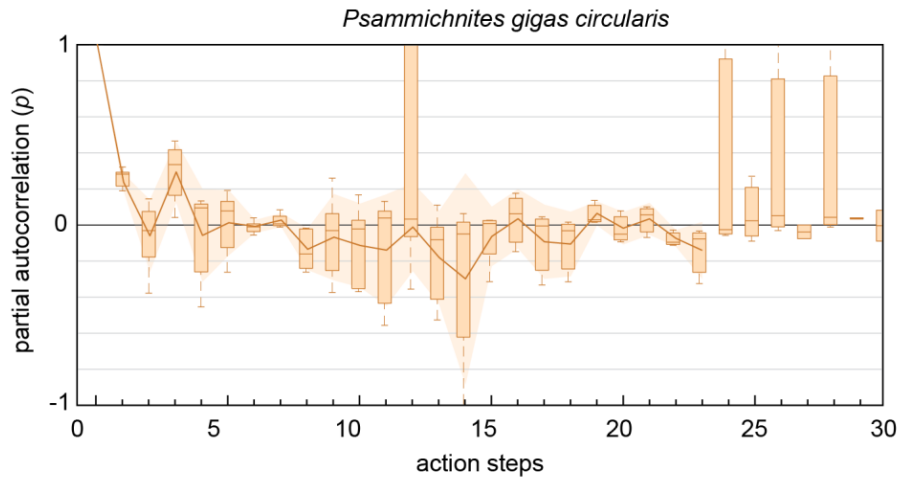

Figure 4.19. Partial autocorrelation functions for *Psammichnites gigas circularis*, performed on turning angles collected at "action step" ($s = 0.7\ w$), determined by the sampling-induced autocorrelation distance (red dashed line in Fig. 3). Black lines denote mean $p$ values, while translucent fill indicates the standard error of the mean ($SE = \frac{\sigma}{\sqrt{n}}$).